\documentstyle[aps,a4]{revtex}
\begin{document}
\pagestyle{myheadings}\markright{Macroconstraints from
Microsymmetries of Macrosystems}%
\draft%
\title{Macroconstraints from Microsymmetries of Macrosystems}
\author{Ludwik Turko\thanks{E-mail: turko@ift.uni.wroc.pl}}
\address{Institute of Theoretical Physics, University of
Wroc\l aw,\\
 pl. Maksa Borna 9, 50-204 Wroc\l aw, Poland}
\author{Jan Rafelski\thanks{E-mail: rafelski@physics.arizona.edu}}
\address{Department of Physics, University of Arizona,\\
 Tucson, AZ 85721, U.S.A.}
\date{November 13, 2000}
\maketitle
\begin{abstract}
The dynamics governing the evolution of a many body system is
constrained by a nonabelian local symmetry. We obtain a general
form of the global macroscopic conditions assuring that at the
microscopic level  the evolution respects the overall symmetry
constraint. These conditions does not depend on detailed dynamics
of the system. It is shown that a nonabelian internal symmetry
leads to the new subsidiary condition imposed on the system. A
structure of constraints is analysed.
\end{abstract}

The statistical mechanics of finite systems is a powerful tool to
describe high-energy strong interaction phenomena. There is a well
established tradition of this approach, starting from classical
ideas of Koppe\cite{Koppe} and Fermi\cite{Fermi}. To get a better
insight into properties of systems under considerations even the
very refined relativistic statistical
thermodynamics\cite{Hagedorn} must be supplemented with a kinetic
theory\cite{Rafelski82,Koch86,Matsui86,Biro93}. An available phase
space is always constrained by internal symmetries of the system.
A consistent formulation of thermodynamics for with internal
symmetries taken into account\cite{TurRed,Turko} allowed for
suitable modifications of the equilibrium
distributions\cite{Turko,RafDan,MullRaf}. In the case of the
kinetic theory the situation is more involved. The influence of
the internal symmetry on the generalized Vlasov - Boltzmann
kinetic equations was considered recently by the
authors\cite{TurRaf}.

In this paper we consider a more general scheme which allows to
obtain a set of conditions due to the internal symmetry preserved
by the underlying local interaction. We obtain a generic form for
these conditions, independent on details of the kinetic equations.

Let us consider a multiparticle quantum system with the local
interaction invariant with respect to the internal symmetry group
$G$. We shall call this underlying symmetry of microscopic
interactions a \underline{microsymmetry} of the system. The system
transforms under a given representation of the symmetry group. We
shall call this property a \underline{macrosymmetry} of the
system. Let us try to answer a question: is macrosymmetry
preserved during a time evolution of the system?

The system consists of particles belonging to multiplets
$\alpha_j$ of the symmetry group. Particles from the given
multiplet $\alpha_j$ are characterized by quantum numbers $\nu_j$
--- related to the symmetry group, and quantum numbers $\zeta_j$
characterizing different multiplets of the same representation
$\alpha_j$. The number of particles species is:
\begin{equation}\label{species}
W=\sum_j\sum_{\zeta_j}d(\alpha_j)\, .
\end{equation}
where $d(\alpha)$ is the dimension of the representation $\alpha.$

The number of particles of the specie
$\{\alpha,\nu_\alpha;\zeta\}$ is denoted here by
$N^{(\alpha)}_{\nu_\alpha;(\zeta)}.$ These occupation numbers are
time dependent until the system reaches the chemical equilibrium.
However the representation of the symmetry group for the system
remains constant in the course of a time evolution.

Let us consider a multiparticle state vector in occupation number
representation: $\left\vert N^{(\alpha_1)}_{\nu_{\alpha_1}},\dots,
N^{(\alpha_n)}_{\nu_{\alpha_n}}\right\rangle .$ All other
variables, related to phase-space properties of the system are
suppressed here. This vector describes symmetry properties of our
systems and transforms as a direct product representation of the
symmetry group $G$. This representation is of the form:
\begin{equation}\label{prod}
\alpha_1^{N^{(\alpha_1)}} \otimes \alpha_2^{N^{(\alpha_2)}}
\otimes \cdots \otimes \alpha_n^{N^{(\alpha_n)}}.
\end{equation}
A multiplicity $N^{(\alpha_j)}$ of the representation $\alpha_j$
in this product is equal to a number of particles which transform
under this representation:
\begin{equation}
N^{(\alpha_j)} = \sum_j\left(\sum_{\zeta_j}\,
N^{(\alpha_j)}_{\nu_{\alpha_j};(\zeta_j)}\right)= \sum_j\,
N^{(\alpha_j)}_{\nu_{\alpha_j}}\,.
\end{equation}
The representation given by Eq. (\ref{prod}) can be decomposed
into direct sum of irreducible representations $\Lambda_k$.
Corresponding states are denoted as $\left\vert \Lambda_k,
\lambda_{\Lambda_k}; {\mathcal N}\right\rangle$ where
$\lambda_{\Lambda_k}$ is an index numbering members of the
representation $\Lambda$ and ${\mathcal N}$ is a total number of
particles
\begin{equation}\label{number}
{\mathcal N}=\sum_k\, N^{(\alpha_k)}_{\nu_{\alpha_k}}\,.
\end{equation}
Each physical state can be decomposed into irreducible
representation base states $ \left\vert \Lambda_k,
\lambda_{\Lambda_k}; {{\mathcal N};\xi_{\Lambda_k}}\right\rangle
$. Variables $\xi_{\Lambda}$ are degeneracy parameters required
for the full description of a state in the "symmetry space". Let
us consider a projection operator ${\mathcal P}^{\Lambda}$ on the
subspace spanned by all states transforming under representation
$\Lambda$.

\begin{eqnarray}\label{proj1states}
{\mathcal P}^{\Lambda}\left\vert
N^{(\alpha_1)}_{\nu_{\alpha_1}},\dots,
N^{(\alpha_n)}_{\nu_{\alpha_n}}\right\rangle=
\sum_{\xi_{\Lambda}}\!^{\oplus} \left\vert \Lambda,
\lambda_{\Lambda};\xi_\Lambda\right\rangle{\mathcal
C}^{\Lambda,\lambda_{\Lambda}}_{\{N^{(\alpha_1)}_{\nu_{\alpha_1}},
\dots,\,N^{(\alpha_n)}_{\nu_{\alpha_n}}\}}(\xi_\Lambda)\,.
\end{eqnarray}
This operator has the generic form (see e.g. \cite{Wigner}):
\begin{equation}\label{proj1}
{\mathcal
P}^{\Lambda}=d(\Lambda)\int\limits_G\,d\mu(g)\bar\chi^{(\Lambda)}(g)U(g)\,.
\end{equation}
Here $\chi^{(\Lambda)}$ is the character of the representation
$\Lambda$, $d\mu(g)$ is the invariant Haar measure on the group,
and $U(g)$ is an operator transforming a state under
consideration. We will use the matrix representation:
\begin{eqnarray}\label{transf}
&&U(g)\left\vert N^{(\alpha_1)}_{\nu_{\alpha_1}},\dots,
N^{(\alpha_n)}_{\nu_{\alpha_n}}\right\rangle\\
&&=\sum\limits_{\nu_1^{(1)},\dots,\nu_n^{(N_{\nu_n})}}\,
D^{(\alpha_1)}_{\nu_1^{(1)}\nu_1}\!\!\cdots
D^{(\alpha_1)}_{\nu_1^{(N_{\nu_1})}\nu_1}\!\!\cdots
D^{(\alpha_n)}_{\nu_n^{(1)}\nu_n}\cdots
D^{(\alpha_n)}_{\nu_n^{(N_{\nu_n})}\nu_n} \left\vert
N^{(\alpha_1)}_{\nu_{\alpha_1}},\dots,
N^{(\alpha_n)}_{\nu_{\alpha_n}}\right\rangle\,.\nonumber
\end{eqnarray}
$D^{(\alpha_n)}_{\nu,\nu}$ is a matrix elements of the group
element $g$ corresponding to the representation $\alpha$. Notation
convention in Eq.\,(\ref{transf}) arises since  there are
$N^{(\alpha_j)}_{\nu_{\alpha_j}}$ states transforming under
representation $\alpha_j$ and having quantum numbers of the
$\nu_{\alpha_j}$-th member of a given multiplet.

Let
$\overline{P^{\Lambda,\lambda_{\Lambda}}_{\{N^{(\alpha_1)}_{\nu_{\alpha_1}},
\dots,\,N^{(\alpha_n)}_{\nu_{\alpha_n}}\}}}$ denotes the
probability that $N^{(\alpha_1)}_{\nu_{\alpha_1}},\dots,
N^{(\alpha_n)}_{ \nu_{\alpha_n}}$ particles transforming under the
symmetry group representations $\alpha_1,\dots,\alpha_n$ combine
into $\mathcal N$ particle state transforming under representation
$\Lambda$ of the symmetry group. This is equal to a norm of the
vector (\ref{proj1states}) and it is given by
\begin{eqnarray}\label{normP}
\qquad\left\langle N^{(\alpha_1)}_{\nu_{\alpha_1}},
\cdots,N^{(\alpha_n)}_{\nu_{\alpha_n}}\right\vert {\mathcal P
}^{\Lambda}\left\vert N^{(\alpha_1)}_{\nu_{\alpha_1}},\dots,
N^{(\alpha_n)}_{\nu_{\alpha_n}}\right\rangle=
\sum\limits_{\xi_\Lambda}\vert{\mathcal
C}^{\Lambda,\lambda_{\Lambda}}_{\{N^{(\alpha_1)}_{\nu_{\alpha_1}},
\dots,\,N^{(\alpha_n)}_{\nu_{\alpha_n}}\}}(\xi_\Lambda)\vert^2\,.
\end{eqnarray}
Left hand side of this equation can be calculated directly from
Eqs.(\ref{proj1}) and (\ref{transf}). One gets finally
\begin{equation} \label{weights}
\overline{P^{\Lambda,\lambda_{\Lambda}}_{\{N^{(\alpha_1)}_{\nu_{\alpha_1}},
\dots,\,N^{(\alpha_n)}_{\nu_{\alpha_n}}\}}}={\mathcal
A}^{\{{\mathcal N}\}}
d(\Lambda)\int\limits_G\,d\mu(g)\bar\chi^{(\Lambda)}(g)
[D^{(\alpha_1)}_{\nu_1\nu_1}]^{N^{(\alpha_1)}_{\nu_{\alpha_1}}}
\cdots
[D^{(\alpha_n)}_{\nu_n\nu_n}]^{N^{(\alpha_n)}_{\nu_{\alpha_n}}}\,.
\end{equation}
where ${\mathcal A}^{\{{\mathcal N}\}}$ is a permutation
normalization factor. For particles of the kind $\{\alpha,\zeta\}$
we included in Eq.\,(\ref{weights}) the permutation factor:
\begin{equation}\label{permfac1}
{\mathcal A}^\alpha_{(\zeta)}=
\frac{N^{(\alpha)}_{(\zeta)}!}{\prod\limits_{\nu_\alpha}
N^{(\alpha)}_{\nu_\alpha;(\zeta)}!}\,.
\end{equation}
The permutation factor ${\mathcal A}^{\{{\mathcal N}\}}$ is a
product of all "partial" factors
\begin{equation}\label{permfactor}
{\mathcal A}^{\{{\mathcal N}\}}=
\prod\limits_j\prod_{\zeta_j}{\mathcal
A}^{\alpha_j}_{(\zeta_j)}\,.
\end{equation}
Because of macrosymmetry all weights in Eq.\,(\ref{weights})
should be constant in time. So we have
\begin{equation} \label{cond}
\frac{d}{dt}\overline{P^{\Lambda,\lambda_{\Lambda}}_{\{N^{(\alpha_1)}_{\nu_{\alpha_1}},
\dots,\,N^{(\alpha_n)}_{\nu_{\alpha_n}}\}}}=0\,.
\end{equation}
This condition assures that in a dynamical evolution the
macrosymmetry of the system is preserved. By mean of Eq.
(\ref{weights}) the macrosymmetry conservation condition
(\ref{cond}) is converted into a general time evolution condition
of the form
\begin{eqnarray}\label{deriv}
0 &=& \frac{d\,{\mathcal A}^{\{{\mathcal
N}\}}}{dt}d(\Lambda)\int\limits_G d\mu(g)\bar\chi^{(\Lambda)}(g)
[D^{(\alpha_1)}_{\nu_1\nu_1}]^{N^{(\alpha_1)}_{\nu_{\alpha_1}}}\cdots
[D^{(\alpha_n)}_{\nu_n\nu_n}]^{N^{(\alpha_n)}_{\nu_{\alpha_n}}}\\
&&+\sum_{j=1}^n\sum_{\nu_{\alpha_j}}\,
\frac{d\,N^{(\alpha_j)}_{\nu_{\alpha_j}}}{dt} {\mathcal
A}^{\{{\mathcal N}\}}d(\Lambda)\int\limits_G
d\mu(g)\bar\chi^{(\Lambda)}(g)
[D^{(\alpha_1)}_{\nu_1\nu_1}]^{N^{(\alpha_1)}_{\nu_{\alpha_1}}}
\cdots
[D^{(\alpha_n)}_{\nu_n\nu_n}]^{N^{(\alpha_n)}_{\nu_{\alpha_n}}}
\log[D^{(\alpha_j)}_{\nu_j\nu_j}]\,.\nonumber
 \end{eqnarray}
All integrals which appear in Eq.\,(\ref{weights}) and
Eq.\,(\ref{deriv}) can be expressed explicitly in an analytic form
for any compact symmetry group.

To write an expression for the time derivative of the
normalization factor ${\mathcal A}^{\{{\mathcal N}\}}$ we perform
analytic continuation from integer to continuous values of
variables $N^{(\alpha_n)}_{\nu_{\alpha_n}}.$ Thus we replace all
factorials by the gamma functions of corresponding arguments. We
encounter here also the digamma function $\psi$ \cite{Abram}:
\begin{equation}\label{digamma}
\psi(x)=\frac{d\, \log\Gamma(x)}{d\,x}\,.
\end{equation}
This allows to write:
\begin{equation}\label{evfactor}
\frac{d\,{\mathcal A}^{\{{\mathcal N}\}}}{dt} = {\mathcal
A}^{\{{\mathcal N}\}}\sum_j\sum_{\zeta_j}
 \left[\frac{d\,N^{(\alpha_j)}_{(\zeta_j)}}{dt}
 \psi(N^{(\alpha_j)}_{(\zeta_j)}+1)
 -\sum_{\nu_{\alpha_j}}
\frac{d\,N^{(\alpha_j)}_{\nu_{\alpha_j};(\zeta_j)}}{dt}
\psi(N^{(\alpha_j)}_{\nu_{\alpha_j};(\zeta_j)}+1)\right]\,.
\end{equation}
These subsidiary conditions (\ref{deriv}) and (\ref{evfactor}) are
the necessary conditions for preserving the internal symmetry on
the macroscopic level. Rates of change
${d\,N^{(\alpha)}_{\nu_\alpha;(\zeta)}}/{dt}$ are related to
``macrocurrents", which are counterparts of ``microcurrents"
related directly to a symmetry on a microscopic level via the
Noether theorem. Eqs. (\ref{deriv}) and (\ref{evfactor}) can be
considered as a set of conditions on macrocurrents to provide
consistency with the overall symmetry of the system.

For multiplicities encountered in high energy heavy ion collisions
processes all digamma functions in Eq. (\ref{evfactor}) can be
replaced by corresponding logarithmic functions according to the
asymptotic formula\cite{Abram}:
\begin{equation}\label{asymptdigamma}
\psi(N+1)\approx\log N
\end{equation}
This gives a simplified form of Eq. (\ref{evfactor}):
\begin{equation}\label{evfactorlog}
\frac{d\,{\mathcal A}^{\{{\mathcal N}\}}}{dt} = {\mathcal
A}^{\{{\mathcal N}\}}\sum_j\sum_{\zeta_j}
 \left[\frac{d\,N^{(\alpha_j)}_{(\zeta_j)}}{dt}
 \log N^{(\alpha_j)}_{(\zeta_j)}
 -\sum_{\nu_{\alpha_j}}
\frac{d\,N^{(\alpha_j)}_{\nu_{\alpha_j};(\zeta_j)}}{dt} \log
N^{(\alpha_j)}_{\nu_{\alpha_j};(\zeta_j)}\right]\,.
\end{equation}
One obtains then from Eq.(\ref{deriv})
\begin{eqnarray}\label{derivasympt}
&&\sum_{j=1}^n\sum_{\nu_{\alpha_j}}\,
\frac{d\,N^{(\alpha_j)}_{\nu_{\alpha_j}}}{dt} \frac{d\,\log
\widetilde {\cal
P}^{\Lambda,\lambda_{\Lambda}}_{\{N^{(\alpha_1)}_{\nu_{\alpha_1}},
\dots,\,N^{(\alpha_n)}_{\nu_{\alpha_n}}\}}}{d\,N^{(\alpha_j)}_{\nu_{\alpha_j}}}\\
&=&\sum_j\sum_{\zeta_j}
\left(-\frac{d\,N^{(\alpha_j)}_{(\zeta_j)}}{dt} \log
N^{(\alpha_j)}_{(\zeta_j)}+\sum_{\nu_{\alpha_j}}
\frac{d\,N^{(\alpha_j)}_{\nu_{\alpha_j};(\zeta_j)}}{dt} \log
N^{(\alpha_j)}_{\nu_{\alpha_j};(\zeta_j)}\right)\nonumber \, .
\end{eqnarray}
where
\begin{equation}\label{calP}
\widetilde {\cal
P}^{\Lambda,\lambda_{\Lambda}}_{\{N^{(\alpha_1)}_{\nu_{\alpha_1}},
\dots,\,N^{(\alpha_n)}_{\nu_{\alpha_n}}\}}=\int\limits_G
d\mu(g)\bar\chi^{(\Lambda)}(g)
[D^{(\alpha_1)}_{\nu_1\nu_1}]^{N^{(\alpha_1)}_{\nu_{\alpha_1}}}
\cdots
[D^{(\alpha_n)}_{\nu_n\nu_n}]^{N^{(\alpha_n)}_{\nu_{\alpha_n}}}
\end{equation}
is analytically extended for continuous values of variables
$N^{(\alpha_j)}_{\nu_{\alpha_j}}.$

Eqs (\ref{deriv}) and (\ref{derivasympt}) are meaningful only for
nonzero values of coefficients (\ref{calP}). Let our symmetry
group be a Lie group of rank $k$. It means physically that there
are exactly $k$ simultaneously measurable charges related to the
group $G$. It is easy to see that weights (\ref{weights}) are
nonzero only for such values of $N^{(\alpha_j)}_{\nu_{\alpha_j}}$
which are consistent with conservation of those charges. For the
isospin $SU(2)$ group that is the third component of the isospin,
for the flavour $SU(3)$ that would be the third component of the
isospin and the hypercharge. In general case one has $k$ linear
relations between variables $N^{(\alpha_j)}_{\nu_{\alpha_j}}$ what
reduces correspondingly the number of independent variables.

New results are obtained only for nonabelian symmetries. The local
abelian charge conservation on the microscopic level is equivalent
to the charge conservation of the multiparticle system. In the
case of the abelian symmetry group all representations are
one-dimensional ones, labelled by the charge value $\alpha$. All
related permutation factors (\ref{permfac1}) are equal to $1$ and
Eq.(\ref{weights}) gives
\begin{equation} \label{abelweights}
\overline{P^{\Lambda}_{\{N^{(\alpha_1)},
\dots,\,N^{(\alpha_n)}\}}} =\frac{1}{2\pi}\int\limits_{0}^{2\pi}
d\varphi\, e^{-i\Lambda\varphi}e^{i\alpha_1 N^{(\alpha_1)}}\cdots
e^{i\alpha_n N^{(\alpha_n)}}=\delta_{\Lambda,\, \alpha_1
N^{(\alpha_1)} + \cdots + \alpha_n N^{(\alpha_n)}}\,.
\end{equation}
We see that for abelian internal symmetries Eq.(\ref{deriv}) gives
an obvious result
\begin{equation} \label{abel}
\alpha_1\frac{d N^{(\alpha_1)}}{dt} + \cdots + \alpha_n\frac{d
N^{(\alpha_n)}}{dt}=0\,.
\end{equation}
For nonabelian symmetries one gets similar relations for charges
belonging to the center of the group $G$.

The ``nonabelian" condition (\ref{deriv}) is much more involved.
The time derivatives ${d\,N^{(\alpha)}_{\nu_\alpha;(\zeta)}}/{dt}$
can be obtained from analysis of the microscopic basic
interaction. The most general approach \cite{RafTur: prep} goes
through Heisenberg equation
\begin{equation}\label{Heis}
\frac{d\,\hat N}{dt}= i[\hat H,\hat N]
\end{equation}
averaged over states $\left\vert
N^{(\alpha_1)}_{\nu_{\alpha_1}},\dots,
N^{(\alpha_n)}_{\nu_{\alpha_n}}\right\rangle .$ $\hat N$ denotes
here a particle number operator.

Another way is to use semiclassical kinetic equations for
distribution functions $f^{(\alpha_i,\nu_i)}_{(\zeta)}(\Gamma,\vec
r,t)$. The variables $(\Gamma,\vec r)$ denote a  set of phase -
space variables. The number of particles of the specie
$\{\alpha,\nu_\alpha;\zeta\}$ is:
\begin{equation}
N^{(\alpha)}_{\nu_\alpha;(\zeta)}(t)=\int\,dV d\Gamma
f^{(\alpha,\nu_\alpha)}_{(\zeta)}(\Gamma,\vec r,t)\,.
\end{equation}
The case of the generalized Vlasov - Boltzmann kinetic equations
was considered in \cite{TurRaf}.

In general the derivatives
${d\,N^{(\alpha)}_{\nu_\alpha;(\zeta)}}/{dt}$ are expressed as
functionals over variables $N^{(\alpha_j)}_{\nu_{\alpha_j}}:$
\begin{equation}\label{functional}
\frac{d\,N^{(\alpha)}_{\nu_\alpha;(\zeta)}}{dt}={\mathcal
F}^{(\alpha)}_{\nu_\alpha;(\zeta)}
\left[N^{(\alpha_1)}_{\nu_{\alpha_1}},\dots,N^{(\alpha_n)}_{\nu_{\alpha_n}}\right]\,
.
\end{equation}
Substituting this into Eq. (\ref{deriv}) or Eq.
(\ref{derivasympt}) one gets a new relation between variables
$N^{(\alpha_j)}_{\nu_{\alpha_j}}.$ This leads in principle to the
further reduction of the number of independent variables. Because
of a highly nonlinear character of the resulting constraint it
would be too optimistic to expect an analytic solution of this
constraint. However, it is possible to get at least numerical
estimations\cite{RafTur: prep} based on kinetic equations
(\ref{functional}) and taking into account $k$ solvable
constraints related to the center of the group $G$ and the
implicit constraint resulting from Eq. (\ref{deriv}) or Eq.
(\ref{derivasympt}).

There is a simple geometrical interpretation of our problem.
Solutions of Eqs (\ref{functional}) span $W$-dimensional manifold
parameterized by the time variable, with $W$ given by Eq.
(\ref{species}). Constraints reduce this manifold to the
$(W-k-1)$-dimensional submanifold of physical states consistent
with the underlying internal symmetry.

New constraints lead to decreasing number of available states for
the system during its time evolution. The equilibrium distribution
can be constructed by the Lagrange multipliers method. The
multipliers related to the ``central" constraints are well known
chemical potentials. The multiplier related to the ``nonabelian"
constraint (\ref{deriv}) is more complicated. Because the
constraint is a nonlinear one, the corresponding multiplier cannot
be treated as a standard additive thermodynamical potential.

\acknowledgments{Work supported in part by the Polish Committee
for Scientific Research under contract KBN-2~P03B~030~18 and by a
grant from the U.S. Department of Energy, DE-FG03-95ER40937\,.}


\begin{references}

\bibitem{Koppe} H.~Koppe, {\it Zs.f.Naturforschung} {\bf 3a},
251, (1948),\\
H. Koppe, {\it Phys. Rev.} {\bf 76}, 688, (1949)\,.

\bibitem{Fermi} E.~Fermi, {\it Progr. Theor. Phys.} {\bf 5},
570, (1950)\,

\bibitem{Hagedorn} R.~Hagedorn, CERN Yellow Report 71-12 (1971)

\bibitem{Rafelski82}
J.~Rafelski and B.~M\"uller,  {\it Phys. Rev. Lett.}  {\bf 48},
1066, (1982)\,.

\bibitem{Koch86}
P.~Koch, B.~M\"uller and J.~Rafelski,  {\it Phys. Rept.} {\bf
142}, 167, (1986)\,.

\bibitem{Matsui86}
T.~Matsui, B.~Svetitsky and L.D.~McLerran, {\it Phys. Rev.} {\bf
D34}, 783, (1986)\,.

\bibitem{Biro93}
T.S.~Biro, E.~van Doorn, B.~M\"uller, M.H.~Thomma and X.-N.~Wang,
{\it Phys. Rev.} {\bf C48}, 1275, (1993)\,.

\bibitem{TurRed}
K. Redlich and L. Turko, {\it Z. Phys.} {\bf C5}, 201,  (1980)\,.

\bibitem{Turko}
 L. Turko, {\it Phys. Lett.} B {\bf 104}, 153,  (1981)\,.

\bibitem{RafDan}
J. Rafelski and M. Danos,  {\it Phys. Lett.} B {\bf 97}, 279,
(1980)\,.

\bibitem{MullRaf}
B. M\"uller and J. Rafelski,  {\it Phys. Lett.} B {\bf 116}, 274,
(1982)\,.

\bibitem{TurRaf} L. Turko and J. Rafelski, {\it Dynamics of
Multiparticle Systems with non -- Abelian Symmetry}, {\bf
hep-th/0003079}; to be published in {\bf EPJ C}\,.

\bibitem{Wigner}
E. P. Wigner, {\it Group Theory and Its Application to the Quantum
Mechanics of Atomic Spectra}, (Academic Press, New York and
London, 1959)\,.
\bibitem{Abram} M. Abramowitz and I. A. Stegun (eds.), {\it Handbook of
Mathematical Functions}, (National Bureau of Standards, Applied
Mathematics Series $\cdot$  55, 1964)\,.

\bibitem{RafTur: prep}
J. Rafelski and L. Turko --- {\it in preparation}\,.


\end{references}
\end{document}